\documentclass{article}
\usepackage{amsmath}
\usepackage{amsfonts}
\usepackage{amssymb}
\usepackage{graphicx}

\begin{document}

\title{Finite-size effects in one-dimensional Bose-Einstein condensation of photons}
\author{Zhi-Jie Liu and Mi Xie\thanks{Email: xiemi@tju.edu.cn}\\{\footnotesize Department of Physics, School of Science, Tianjin University,
Tianjin 300072, P. R. China}}

\date{}
\maketitle

\begin{abstract}
The Bose-Einstein condensation (BEC) of photons has been realized
in one- and two-dimensional systems. When considering the influence of
finite-size effect, the condensation in the one-dimensional fibre is of
special interest since such a condensation cannot occur in the thermodynamic
limit due to the linear dispersion relation of photons. The finite-size
effect must play a key role in this system and needs a detailed description.
However, the previous theoretical analysis of finite-size effect is often
not accurate enough and only gives the leading-order contribution due to a
divergence difficulty. In this paper, by using an analytical continuation
method to overcome the divergence difficulty, we give an analytical
treatment for the finite-size effect in BEC. The analytical expressions of
the critical temperature or critical particle number with higher order
correction and the chemical potential below the transition point are
presented. Our result shows that in a recent experiment, the deviation
between experiment and theory is overestimated, most of which is caused by
the inaccurate theoretical treatment of the finite-size effect. By taking
into account the next-to-leading correction, we find that the actual
deviation is much smaller.

\end{abstract}

%\keywords{Bose-Einstein condensation, photon condensation, finite-size effect,
% heat kernel expansion}

\section{Introduction}

The Bose-Einstein condensation (BEC) of photons was generally believed to be
impossible since the number of photons is not conserved and the extremely
weak interaction between photons cannot thermalize the gas. However, the
situation changed in recent years. By trapping photons in a dye-filled
microcavity, the BEC has been realized in two-dimensional systems \cite%
{KSVW,SDDVK,SDDWVKW,DSL}. In these experiments, the photons are trapped
between two curved mirrors. The fixed longitudinal momentum gives an
effective mass to the photon and a nonvanishing chemical potential to the
photon gas. The repeated absorbtion and emission cycle of the dye molecules
thermalizes the photon gas. Recently, a one-dimensional photon condensation
is also reported \cite{WBLF}. In the experiment, the photons are confined in
a closed erbium--ytterbium co-doped fiber with a cutoff wavelength. The
existence of the cutoff wavelength gives the photons a nonvanishing chemical
potential.

In the experiments of the photon condensation, the finite particle number
makes the behavior of the phase transition different from the thermodynamic
limit case. In particular, the finite-size effect in the one-dimensional
condensation has some different feature since it cannot occur in
thermodynamic limit due to the linear dispersion relation of photons. This
implies that the finite-size effect has a significant influence in this case
and needs to be carefully analyzed.

Many studies have been devoted to the finite-size effect in BEC. The
corrections to the critical temperature and the condensate fraction have
been found, but often only the leading order term can be obtained \cite%
{KVD,Mullin,Yukalov}. The main obstacle for accurately studying the
finite-size effect is the divergence problem: When taking account into the
contribution from the discrete energy levels of the trapped particles
accurately, most terms in the expressions of thermodynamic quantities become
divergent at the transition point.

To overcome the divergence difficulty, we will use an analytical
continuation method to give an analytical treatment to the problem of photon
condensation \cite{X2018,X2019}. By this way, we will obtain more accurate
expressions of critical temperature and condensate fraction with
next-to-leading corrections. We will also give the analytical expression of
the chemical potential, which is hard to obtain before. At low temperature,
the chemical potential is linear in temperature, which is quite different
from the thermodynamic limit case. The comparison with the numerical
solution confirms our result.

In the experiment of one-dimensional photon condensation \cite{WBLF}, the
deviation of the critical particle number between experiment and theory is
about $5.6\%$. However, according to our result, this deviation is mainly
caused by the inaccurate estimate of the finite-size effect in the previous
studies. Our result shows that the contribution of the finite-size effect
should be $4.2\%$ higher than the previous prediction. Therefore the
agreement of experiment and theory is actually very well.

This paper is organized as follows. In section 2 we give an analytical
treatment to the finite-size effect of the photon condensation in one
dimension. In section 3 we compare our result with experiment. Conclusions
and some discussion are presented in section 4.

\bigskip

\section{Critical temperature and chemical potential}

Consider photons in a one-dimensional closed fiber with length $L$ and index
of refraction $n$. The possible frequencies of the photons are restricted by
periodic boundary conditions as%
\begin{equation}
\omega=m^{\prime}\frac{2\pi c}{nL}\equiv m^{\prime}\Delta,   \label{omega}
\end{equation}
where $m^{\prime}$ is an integer, and we have introduced $\Delta\equiv \frac{%
2\pi c}{nL}$ with $c$ the speed of light in vacuum. If there is a cutoff
frequency $\omega_{0}=m_{0}\Delta$, namely, only the photons with a
frequency higher than $\omega_{0}$ can exist in the fiber and the quantum
number $m^{\prime}$ in equation (\ref{omega}) must be larger than $m_{0}$.
For convenience, we shift the energy spectrum to make the ground-state
energy vanish. Then the spectrum of the photons in the fiber becomes%
\begin{equation}
\varepsilon_{m}=m\hbar\Delta,\text{ \ \ }\left( m=0,1,2,3,\cdots\right)
\label{Em}
\end{equation}
Clearly, the photon in such a system has the same energy spectrum as that of
nonrelativistic particles in a one-dimensional harmonic trap, so these two
kinds of systems should show the same transition behavior.

As we know, the BEC occurs when the number of excited particles $N_{e}$
equals the total number of particles $N$ at the chemical potential $\mu=0$.
The excited photon number is%
\begin{equation}
N_{e}=\sum_{m=1}^{\infty}\frac{1}{e^{\beta\left( \varepsilon_{m}-\mu\right)
}-1},   \label{Ne0}
\end{equation}
where $\beta=1/k_{B}T$ with $k_{B}$ is the Boltzmann constant. In the
thermodynamic limit, the energy spectrum becomes continuous and the density
of states is $\rho\left( \varepsilon\right) =1/\hbar\Delta$, the summation
in equation (\ref{Ne0}) is converted to an integral as%
\begin{equation}
N_{e}=\frac{1}{\hbar\Delta}\int_{0}^{\infty}\frac{1}{e^{\beta\left(
\varepsilon-\mu\right) }-1}d\varepsilon=\frac{1}{\beta\hbar\Delta}%
g_{1}\left( e^{\beta\mu}\right) ,   \label{Ne_TL}
\end{equation}
where $g_{\sigma}\left( z\right)
=\sum_{\ell=1}^{\infty}z^{\ell}/\ell^{\sigma}$ is the Bose-Einstein
integral, which has the following asymptotic behavior%
\begin{equation}
g_{\sigma}\left( e^{\beta\mu}\right) \approx\left\{
\begin{array}{lll}
\zeta\left( \sigma\right) , & \left( \sigma>1\right) &  \\
-\ln\left( -\beta\mu\right) , & \left( \sigma=1\right) &  \\
\Gamma\left( -\sigma+1\right) \frac{1}{\left( -\beta\mu\right) ^{-\sigma+1}},%
\text{ } & \left( \sigma<1\right) & \text{ \ \ \ }\left(
\mu\rightarrow0\right)%
\end{array}
\right.   \label{BEint}
\end{equation}
Then in the thermodynamic limit, $N_{e}$ in equation (\ref{Ne_TL}) is
divergent at $\mu=0$, so there is no phase transition.

On the other hand, in finite systems, the energy spectrum is discrete and
the first excited energy is not $0$, the summation in equation (\ref{Ne0})
should be convergent and a finite critical temperature can be obtained. In
ref. \cite{WBLF}, the summation is approximately converted to an integral
similar to equation (\ref{Ne_TL}) but the lower limit of the integral is
replaced by the first excited energy $\hbar\Delta$. Then the critical
particle number can be calculated as \cite{WBLF}
\begin{equation}
N_{c}^{\left( 0\right) }=\frac{k_{B}T}{\hbar\Delta}\ln\frac{k_{B}T}{%
\hbar\Delta}.   \label{Nc0}
\end{equation}
In this treatment, the interval between the ground state and the first
excited state is taken into account, but the higher levels are still
regarded as continuous. In fact, many previous studies of BEC in finite
systems along the similar line. The finite-size effect of the BEC in
one-dimensional harmonic trap is also discussed in refs. \cite%
{KVD,Mullin,Yukalov}. Though the treatments have some difference, they all
depended on similar approximations and can only give the leading-order
correction similar to equation (\ref{Nc0}) (may differ by a factor).

Obviously, a more rigorous treatment of equation (\ref{Ne0}) is to preform
the summation directly. To do this, one can expand every term in the
summation as
\begin{equation}
N_{e}=\sum_{m=1}^{\infty}\frac{1}{e^{\beta\left( \varepsilon_{m}-\mu\right)
}-1}=\sum_{m=1}^{\infty}\sum_{\ell=1}^{\infty}\left( e^{-\beta\left(
\varepsilon_{m}-\mu\right) }\right)
^{\ell}=\sum_{\ell=1}^{\infty}e^{\ell\beta\mu}K\left(
\ell\beta\hbar\Delta\right) ,   \label{Ne}
\end{equation}
where%
\begin{equation}
K\left( t\right) =\sum_{m=1}^{\infty}e^{-mt}=\frac{1}{e^{t}-1}   \label{Kt}
\end{equation}
is the global heat kernel \cite{KirstenBK,Vassilevich,Gilkey}. For small $t$%
, the heat kernel (\ref{Kt}) can be expanded as a series of $t$,
\begin{equation}
K\left( t\right) =\sum_{k=0}^{\infty}C_{k}t^{k-1},\text{ \ \ \ }\left(
t\rightarrow0^{+}\right)   \label{Ktexp}
\end{equation}
with the coefficients%
\begin{equation}
C_{0}=1,C_{1}=-\frac{1}{2},C_{2}=\frac{1}{12},C_{3}=0,C_{4}=-\frac{1}{720}%
,\cdots   \label{1.050}
\end{equation}
Substituting the heat kernel expansion (\ref{Ktexp}) into equation (\ref{Ne}%
), we have
\begin{equation}
N_{e}=\sum_{k=0}^{\infty}C_{k}\left( \beta\hbar\Delta\right)
^{k-1}g_{1-k}\left( e^{\beta\mu}\right) .   \label{Ne1}
\end{equation}

A similar treatment can also apply to the grand potential and other
thermodynamical quantities, and these quantities are also expressed as
series of the Bose-Einstein integrals. The higher order correction terms can
describe the effects of boundary, potential, or topology, depending on the
specific systems. This heat kernel expansion approach has been applied to
various problems in statistical physics \cite{KirstenBK,DX2003}. However, a
serious difficulty arises when considering the problem of BEC phase
transition. Due to the asymptotic form of the Bose-Einstein integral
equation (\ref{BEint}), every term in equation (\ref{Ne1}) is divergent at $%
\mu\rightarrow0$, and the divergence becomes more severe in the higher
orders. This divergence problem is the main obstacle for treating the
problem of phase transition in finite systems. As mentioned above, in ref.
\cite{WBLF}, the divergence is avoided by replacing the summation of excited
states by an integral approximately, and only gives the leading-order
correction to the critical temperature. If we want to obtain a more accurate
result, the divergence problem in equation (\ref{Ne1}) must be solved. In
the following, we will use an analytical continuation method \cite%
{X2018,X2019} based on the heat kernel expansion to overcome the divergence
problem.

First, substituting the leading term of the asymptotic expansion of each
Bose-Einstein integral (\ref{BEint}) into equation (\ref{Ne1}) gives%
\begin{equation}
N_{e}=-C_{0}\frac{1}{\beta\hbar\Delta}\ln\left( -\beta\mu\right) +\sum
_{k=1}^{\infty}C_{k}\left( \beta\hbar\Delta\right) ^{k-1}\Gamma\left(
k\right) \frac{1}{\left( -\beta\mu\right) ^{k}}.   \label{Ne2}
\end{equation}
The summation in the second term can be represented by the heat kernel if
the gamma function is replaced by the integral form
\begin{equation}
\Gamma\left( \xi\right) =\int_{0}^{\infty}x^{\xi-1+s}e^{-x}dx,\text{ \ \ \ }%
\left( s\rightarrow0\right)   \label{Gamma}
\end{equation}
where we have introduced a small parameter $s$ which will be taken as $0$ at
the end of the calculation. Then equation (\ref{Ne2}) becomes%
\begin{align}
N_{e} & =-C_{0}\frac{1}{\beta\hbar\Delta}\ln\left( -\beta\mu\right)
+\int_{0}^{\infty}dxe^{-x}x^{s}\frac{1}{-\beta\mu}\left[ K\left( \frac{%
x\hbar\Delta}{-\mu}\right) -C_{0}\frac{-\mu}{x\hbar\Delta}\right]  \notag \\
& =-C_{0}\frac{1}{\beta\hbar\Delta}\ln\left( -\beta\mu\right) +\frac {1}{%
-\beta\mu}\Gamma\left( 1+s\right) \sum_{m=1}^{\infty}\left( 1+\frac{%
m\hbar\Delta}{-\mu}\right) ^{-1-s}-C_{0}\frac{1}{\beta\hbar\Delta }%
\Gamma\left( s\right) .   \label{Ne3}
\end{align}
In the last step, the definition of heat kernel (\ref{Kt}) has been employed
to perform the integral. For $\mu\rightarrow0$, the summation in equation (%
\ref{Ne3}) becomes
\begin{equation}
\sum_{m=1}^{\infty}\left( 1+\frac{m\hbar\Delta}{-\mu}\right)
^{-1-s}\approx\sum_{m=1}^{\infty}\left( \frac{m\hbar\Delta}{-\mu}\right)
^{-1-s}=\frac{\left( -\mu\right) ^{1+s}}{\left( \hbar\Delta\right) ^{1+s}}%
\zeta\left( 1+s\right) ,   \label{sum}
\end{equation}
where $\zeta\left( s\right) =\sum_{n=1}^{\infty}n^{-s}$ is the Riemann $\zeta
$-function.

Now taking the limit $s\rightarrow0$ in equation (\ref{Ne3}), we have%
\begin{align}
N_{e} & \approx-\frac{1}{\beta\hbar\Delta}\ln\left( -\beta\mu\right) +\frac{1%
}{\beta\hbar\Delta}\left( \ln\frac{-\mu}{\hbar\Delta}+\gamma _{E}\right)
\notag \\
& =\frac{1}{\beta\hbar\Delta}\left( \ln\frac{1}{\beta\hbar\Delta}+\gamma
_{E}\right) ,   \label{Ne4}
\end{align}
where the Euler constant $\gamma_{E}=0.577216$. In this result, all the
divergent terms of $s$ and $\mu$ are canceled, and the expression of the
excited-state particle number is completely analytical. Therefore, with the
help of the idea of analytical continuous, the heat kernel expansion is
successfully applied to the phase transition point and the divergence is
eliminated.

From equation (\ref{Ne4}), the critical particle number for a given
temperature $T$ is obviously%
\begin{equation}
N_{c}=\frac{k_{B}T}{\hbar \Delta }\left( \ln \frac{k_{B}T}{\hbar \Delta }%
+\gamma _{E}\right) .  \label{Nc}
\end{equation}%
and the critical temperature for a fixed particle number $N$ is%
\begin{equation}
T_{c}=\frac{\hbar \Delta }{k_{B}}\frac{N}{W\left( Ne^{\gamma _{E}}\right) },
\label{Tc}
\end{equation}%
where $W\left( z\right) $ is the Lambert $W$ function, satisfying $z=W\left(
ze^{z}\right) $. This critical temperature is a little lower than the
previous result corresponding to the critical particle number (\ref{Nc0})%
\begin{equation}
T_{0}=\frac{\hbar \Delta }{k_{B}}\frac{N}{W\left( N\right) }.  \label{T0}
\end{equation}%
According to the asymptotic expansion of the Lambert function $W\left(
x\right) \approx \ln x-\ln \ln x$ for $x\rightarrow \infty $, the critical
temperature can be approximated as%
\begin{equation}
T_{c}\approx \frac{\hbar \Delta }{k_{B}}\frac{N}{\ln N+\gamma _{E}-\ln
\left( \ln N+\gamma _{E}\right) }.  \label{Tc2}
\end{equation}%
We retain the second term in the denominator since for a relative small
particle number, e.g. $N\sim 10^{4}$, $\ln \ln N$ is not much smaller than $%
\ln N$.

The condensate fraction is straightforward from equation (\ref{Ne4}),
\begin{equation}
\frac{N_{0}}{N}=1-\frac{1}{N}\frac{k_{B}T}{\hbar\Delta}\left( \ln\frac {%
k_{B}T}{\hbar\Delta}+\gamma_{E}\right) .   \label{N0}
\end{equation}
This result is not very accurate, especially near the transition point. The
reason is that the chemical potential has been assumed to be vanished below
the transition point in the above calculation. As the temperature tends to
the transition point, the deviation of the chemical potential from $0$
becomes larger and larger. To describe the phase transition more accurately,
we need to find the form of the chemical potential.

The expression of chemical potential $\mu$ can be addressed by the help of
the analytical result of the excited-state number (\ref{Ne4}). For a small
but nonzero chemical potential $\mu$, the ground-state particle number which
is about $1/\left( -\beta\mu\right) $ is not zero at the phase transition
point. Then the total particle number $N$ should contain the contributions
from both the ground state and the excited states:
\begin{equation}
N=\frac{1}{-\beta\mu}+N_{e}.   \label{Nt}
\end{equation}
Here the excited-state particle number $N_{e}$ takes the same form as
equation (\ref{Ne3}), but in the summation (\ref{sum}), an extra term which
is proportional to $\mu$ should be added. Similar to the above procedure, we
can obtain
\begin{equation}
N_{0}=\frac{1}{-\beta\mu}-\zeta\left( 2\right) \frac{-\beta\mu}{\left(
\beta\hbar\Delta\right) ^{2}}-\frac{1}{2}+\frac{1}{2}\frac{-\beta\mu}{%
\beta\hbar\Delta},   \label{N02}
\end{equation}
where $N_{0}$ has been given in equation (\ref{N0}). In the right hand side
of equation (\ref{N02}), the last two terms are small. After neglecting
these two terms, the chemical potential can be solved as
\begin{equation}
\mu=-\frac{\sqrt{6}}{\pi}\hbar\Delta\left[ \sqrt{1+\left( \frac{\sqrt{3}}{%
\sqrt{2}\pi}\frac{T_{0}}{T}\frac{N_{0}}{N/\ln N}\right) ^{2}}-\frac {\sqrt{3}%
}{\sqrt{2}\pi}\frac{T_{0}}{T}\frac{N_{0}}{N/\ln N}\right] .   \label{mu}
\end{equation}
An interesting feature of this result is that at low temperature $T\ll T_{c}$%
, the chemical potential is
\begin{equation}
\mu\approx-\hbar\Delta\frac{1}{\ln N}\frac{T}{T_{0}},\text{ \ \ \ \ }\left(
T\ll T_{c}\right)
\end{equation}
which is linearly related to the temperature. This is different from the
thermodynamic limit result
\begin{equation}
\mu=-k_{B}Te^{-\frac{\hbar\Delta}{k_{B}T}N},\text{ \ \ \ \ }\left( T\ll
T_{c}\right)
\end{equation}
which is exponential small. In figure 1 we compare the chemical potential in
equation (\ref{mu}) with the thermodynamic limit one and the exact numerical
solution, and it confirms the above low-temperature result.

\begin{figure}[ptb]
\begin{center}
\includegraphics[height=3in]{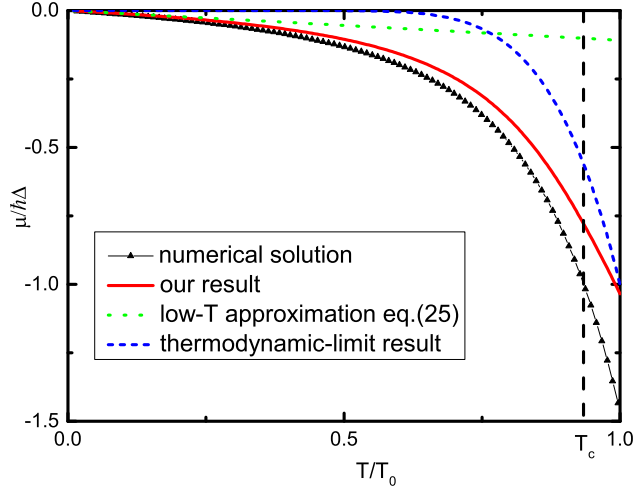}
\end{center}
\caption{The chemical potential below the critical temperature for total
particle number $N=10^4$. Our result of critical temperature $T_c$ is lower
than the previous one $T_0$. At low temperature, the chemical potential is
approximately linear related to the temperature.}
\label{fig1}
\end{figure}

\section{Comparison with the experiment}

In equation (\ref{Nc}), we present the critical particle number of BEC in
the one-dimensional photon system. Compared with the previous result \cite%
{WBLF} given in equation (\ref{Nc0}), the leading-order term is the same,
but our approach also gives a new next-to-leading correction term. This
next-to-leading correction leads to a relative deviation as%
\begin{equation}
\frac{N_{c}-N_{c}^{\left( 0\right) }}{N_{c}^{\left( 0\right) }}=\frac{%
\gamma_{E}}{\ln\frac{k_{B}T}{\hbar\Delta}}\sim\frac{\gamma_{E}}{\ln
N_{c}^{\left( 0\right) }}.
\end{equation}
It indicates that the previous treatment in which the excited states are
regarded as continuous gives a lower prediction of the order of $1/\ln N$,
which is usually not very small in realistic systems.

The relevant experimental parameters are as follows \cite{WBLF}. The length
of the fiber $L=27m$, the refraction coefficient $n=1.444$, the critical
temperature $T=296K$, and the cutoff wavelength $\lambda_{0}=1568nm$. Then
the critical particle numbers given by equations (\ref{Nc0}) and (\ref{Nc})
are%
\begin{equation}
N_{c}^{\left( 0\right) }=1.09\times10^{7},\text{ \ \ }N_{c}=1.14\times
10^{7}.
\end{equation}
Our prediction of $N_{c}$ is about $4.2\%$ higher than $N_{c}^{\left(
0\right) }$ given in Ref. \cite{WBLF}.

In the experiment \cite{WBLF}, the measured quantity is the pump power,
which is proportional to the photon number, and the measurement result is $%
P_{c}^{\exp}=9.5\mu W$. Compared with the theoretical prediction $%
P_{c}^{\left( 0\right) }=9.0\mu W$ \cite{WBLF}, the experimental result is
about $5.6\%$ higher. This is not a large deviation, but according to the
above analysis, most of the deviation is caused by the inaccurate
theoretical prediction. Our result shows that the theoretical value of the
critical particle number should increase by about $4.2\%$. Consequently,
including the next-to-leading contribution of the finite-size effect greatly
improves the agreement between experiment and theory.\

\section{Conclusion and discussion}

In this paper we give a more systematic and accurate discussion on the
finite-size effect in one-dimensional Bose-Einstein condensation of photons.
By using an analytical continuous method based on the heat kernel expansion,
we overcome the divergence difficulty and obtain the next-to-leading order
finite-size corrections on thermodynamic quantities. In the experiment of
one-dimensional photon BEC \cite{WBLF}, the measurement value of the
critical particle number is about $5.6\%$ higher than the previous
theoretical prediction. However, our result shows that the most part of the
deviation arise from the inaccurate analysis of the finite-size effect. When
taking account of the next-to-leading correction, the deviation between
experiment and theory becomes about $1.4\%$.

The magnitude of the finite-size effect is closely related to the spatial
dimension of the system.\ In fact, the most important factor determining the
statistical properties is the density of states, and the density of states
is strongly affected by the spatial dimension. For example, in a
two-dimensional harmonic trap, the leading term of the finite-size
correction to the critical temperature is of the order of $\ln N/\sqrt{N}$,
and the next-to-leading correction has the order of magnitude $1/\sqrt{N}$
\cite{X2019}, which is often negligible. However, in the one-dimensional
case, the leading correction is about $N/\ln N$, and our calculation gives
the next-to-leading term of the order of $1/\ln N$, which is much larger
than the two-dimensional case. In fact, in the thermodynamic limit, photon
BEC cannot occur in one dimension, so the correction of the finite-size
effect in one-dimensional system must be significant. The same behavior also
appears in similar systems, e.g., nonrelativistic particles in
one-dimensional harmonic traps or in two-dimensional boxes.

The method used in this paper is based on the heat kernel expansion. We know
that the heat kernel expansion is a short-wavelength expansion (high-energy
expansion). In principle, it is only applicable to the high temperature and
low density case. When applying the heat kernel expansion to the problem of
phase transition, the divergence problem arises indeed. In this paper,
however, we show that by the help of the analytical continuation method, the
application range of heat kernel expansion can be extended to below the
transition point, and the thermodynamic quantities can also be obtained
analytically.

\end{document}